\documentclass[aps,prl,showpacs,twocolumn,superscriptaddress]{revtex4}
\usepackage{amsmath}
\usepackage{amssymb}
\usepackage{epsfig}
\usepackage{graphicx}
\usepackage{bm}
\begin{document}

\title{Universal behavior of $\rm CePd_{1-x}Rh_x$ Ferromagnet at Quantum Critical Point}
\author{V.R. Shaginyan}\email{vrshag@thd.pnpi.spb.ru}
\affiliation{Petersburg Nuclear Physics Institute, RAS, Gatchina,
188300, Russia}
\author{K.G. Popov}
\affiliation{Komi Science Center, Ural Division, RAS, 3a, Chernova
str. Syktyvkar, 167982, Russia}
\author{S.A. Artamonov} \affiliation{Petersburg Nuclear Physics Institute, RAS, Gatchina,
188300, Russia}

\begin{abstract}
The heavy-fermion metal $\rm CePd_{1-x}Rh_x$ can be tuned from
ferromagnetism at $x=0$ to non-magnetic state at some critical
concentration $x_c$. The non-Fermi liquid behavior  (NFL) at
$x\simeq x_c$ is recognized by power low dependence of the specific
heat $C(T)$ given by the electronic contribution, magnetic
susceptibility $\chi(T)$ and volume expansion coefficient
$\alpha(T)$ at low temperatures:
$C/T\propto\chi(T)\propto\alpha(T)/T\propto1/\sqrt{T}$. We also
demonstrate that the behavior of normalized effective mass $M^*_N$
observed in $\rm CePd_{1-x}Rh_x$ at $x\simeq 0.8$ agrees with that
of $M^*_N$ observed in paramagnetic $\rm CeRu_2Si_2$ and conclude
that these alloys exhibit the universal NFL thermodynamic behavior
at their quantum critical points. We show that the NFL behavior of
$\rm CePd_{1-x}Rh_x$ can be accounted for within frameworks of
quasiparticle picture and fermion condensation quantum phase
transition, while this alloy exhibits a universal thermodynamic NFL
behavior which is independent of the characteristic features of the
given alloy such as its lattice structure, magnetic ground state,
dimension etc.
\end{abstract}
 \maketitle

The nature of the non-Fermi liquid (NFL) behavior observed in
heavy-fermion (HF) metals is still hotly debated. It is widely
believed that the observed behavior is determined by quantum phase
transitions which occur at  quantum critical points, while the
proximity of a system to quantum critical points creates its NFL
behavior brought about by the corresponding thermal and quantum
fluctuations suppressing quasiparticle excitations \cite{voj,loh}.
A quantum critical point (QCP) can arise by suppressing the
transition temperature $T_c$ of a ferromagnetic (or
antiferromagnetic) phase to zero by tuning some control parameters
other than temperature, such as pressure, magnetic field, or doping
$x$ as it takes place in the case of the HF ferromagnet $\rm
CePd_{1-x}Rh_x$ \cite{sereni,pikul} or the HF metal $\rm
CeIn_{3-x}Sn_x$ \cite{kuch}. QCPs are of great interest due to
their singular ability to influence the thermodynamic properties of
materials producing the NFL behavior. The NFL behavior around QCPs
manifests itself in various anomalies. One of them is power in $T$
variations of the specific heat $C(T)$, thermal expansion
$\alpha(T)$, magnetic susceptibility $\chi(T)$ etc.
\cite{voj,loh,ste}.

Measurements on $\rm CePd_{1-x}Rh_x$ show that around concentration
$x=x_c\simeq 0.9$ the suppression of the ferromagnetic phase takes
place, so that this alloy is tuned from ferromagnetism at $x=0$ to
non-magnetic state at QCP with the critical concentration $x_c$
\cite{sereni,pikul}. Studies of the NFL behavior revealed in the HF
metal $\rm CePd_{1-x}Rh_x$ \cite{sereni,pikul} are of great
interest since this alloy is a three dimensional ferromagnet.
Basing on the theory of critical fluctuations which claims that
these are responsible for the corresponding NFL
behavior\cite{voj,loh,ste}, one can assume that the NFL behavior
demonstrating by $\rm CePd_{1-x}Rh_x$ is to be different from that
of $\rm CeNi_2Ge_2$ exhibiting a paramagnetic ground state
\cite{geg1} or from that of the antiferromagnetic cubic HF metal
$\rm CeIn_{3-x}Sn_x$ \cite{kuch}. Obviously the corresponding
critical fluctuations taking place at QCPs in the mentioned three
different HF metals are different, therefore one cannot expect to
observe a universal behavior demonstrating by these metals, while
the traditional theory has no grounds to consider these QCPs as a
single QCP. Moreover, the distinctive features between
ferromagnetic, antiferromagnetic and paramagnetic systems suggest
intrinsic differences in their QCPs resulting in the difference of
their thermodynamic properties, and the theory predicts that
magnetic, thermal and transport properties of these systems have to
be different \cite{voj,loh,ste,sereni,pikul,kir}.

At the critical concentration $x_c$, measurements on $\rm
CePd_{1-x}Rh_x$ show that the specific heat
$C(T)/T\propto1/\sqrt{T}$, while around that concentration $C/T$
and $\chi(T)$ coincide in their temperature dependence,
$C(T)/T\propto\chi(T)\propto1/\sqrt{T}$ \cite{sereni,pikul}.
Moreover, as we shall see it proved to be
$\alpha(T)/T\propto1/\sqrt{T}$ and this NFL behavior of the thermal
expansion coefficient coincides with that of $\alpha(T)$ observed
in the HF metals $\rm CeNi_2Ge_2$  \cite{geg1} and $\rm
CeIn_{3-x}Sn_x$ \cite{sereni,pikul}. The observed power laws and
relationships between them can be hardly accounted for within
scenarios based on the QCP occurrence with quantum and thermal
fluctuations \cite{voj,loh,shag,sereni} when quasiparticles are
suppressed, for there is no reason to expect that $C(T),\,
\chi(T)$, $\alpha(T)$ and other thermodynamic propreties are
affected by fluctuations in a correlated fashion.

These demonstrate that the fluctuations are not responsible for the
observed behavior, and if they are not, what kind of physics
determines the NFL behavior? Fortunately, the direct observations
of quasiparticles in CeCoIn$_5$ have been reported recently
\cite{pag}. On the other hand, when the electronic system of HF
metals undergoes the fermion condensation quantum phase transition
(FCQPT), the fluctuations are strongly suppressed and cannot
destroy the quasiparticles which survive down to lowest
temperatures and we can safely suggest that quasiparticles are
responsible for the NFL behavior observed in HF metals
\cite{shag,shag1,ams,ams1}.

In this letter we show that the NFL behavior of the thermal
expansion coefficient $\alpha(T)/T\propto1/\sqrt{T}$ observed in
$\rm CePd_{1-x}Rh_x$ coincides with that of $\alpha(T)$ observed in
both $\rm CeNi_2Ge_2$ and $\rm CeIn_{3-x}Sn_x$. While the NFL
behavior of the ferromagnet $\rm CePd_{1-x}Rh_x$ related to the
uniform temperature dependence of
$C(T)/T\propto\chi(T)\propto\alpha(T)/T\propto1/\sqrt{T}$ can be
accounted for within the framework of quasiparticle picture and
FCQPT. We demonstrate this alloy is of great interest as it
exhibits the universal NFL thermodynamic behavior at its QCP. This
behavior is independent of the characteristic features of the given
alloy such as its lattice structure, magnetic ground state,
dimension etc. We also conclude that numerous CQPs assumed to be
responsible for the NFL behavior of the thermal expansion
coefficient and other thermodynamic properties observed in HF
metals can be substituted by the only QCP related to FCQPT.

To study the low temperature universal features of HF metals, we
use a model of homogeneous HF liquid with effective mass $M^*$ in
order to avoid the complications associated with the crystalline
anisotropy of solids. This is possible since we consider the
universal behavior related to the power-law divergences of
observable values like the effective mass, thermal expansion
coefficient, specific heat etc. These divergences are determined by
small (as compared to those from unit cell of the corresponding
reciprocal lattice) momenta transfer so that the contribution from
larger momenta can be safely ignored.

To describe the effective mass $M^*$ as a function of temperature
and applied magnetic fields $B$ when the system approaches FCQPT
from the disordered side, $x\to x_{FC}$, we use the Landau equation
connecting the effective mass $M^*(T,B)$ with the bare mass $M$ and
Landau interaction amplitude $F({\bf p}_1,{\bf p}_2,x)$ \cite{land}
\begin{equation}\label{M1}
\frac{1}{M}=\frac{1}{M^*(T,B)}+\int \frac{{\bf
p_F}}{p_F^2}\frac{\partial F({\bf p_F},{\bf p},x)}{\partial {\bf
p_F}}n({\bf p},T,B)\frac{d{\bf p}}{(2\pi)^3},
\end{equation} where $n({\bf p},T,B)$ is the quasiparticle
distribution function \begin{equation}\label{FD}n({\bf p},T,B)=
\left\{1+\exp\left[\frac{(\varepsilon({\bf p},T,B)-\mu(B))}
{T}\right]\right\}^{-1}. \end{equation} Here both the
single-particle energy $\varepsilon({\bf p},T,B)$ and chemical
potential $\mu(T,B)$ depend on temperature and magnetic field. It
follows from Eq. (\ref{FD}) that at $B\to0$ and $T\to0$ the
distribution function $n({\bf p},T,B)\to\theta(p_F-p)$ with
$\theta(p_F-p)$ being the step function and we obtain from Eq.
(\ref{M1}) that \cite{land,pfit,shag_fc}
\begin{equation}\label{M2} M^*(x)=\frac{M}{1-N_0F^1(p_F,p_F,x)/3}
\simeq A+\frac{B}{x-x_{FC}}.\end{equation} Here $N_0$ is the
density of states of a free electron gas, $p_F$ is Fermi momentum,
$F^1(p_F,p_F)$ is the $p$-wave component of Landau interaction
amplitude, $A$ and $B$ are constants. Since Landau Fermi liquid
(LFL) theory implies the number density in the form
$x=p_F^3/3\pi^2$, we can rewrite the amplitude as
$F^1(p_F,p_F,x)=F^1(x)$. When $x\to x_{FC}$, $F^1(x)$ being a
function of $x$ achieves some value at which the denominator tends
to zero so that the effective mass diverges at $T=0$ as seen from
Eq. (\ref{M2}).

At first let us consider the dependence of the effective mass on
temperature. Upon using Eq. (\ref{M2}) and introducing the function
$\delta n({\bf p},T)=n({\bf p},T)-\theta(p_F-p)$, we transform Eq.
(\ref{M1}) and it takes the form
\begin{equation}\label{M3}
    \frac{1}{M^*(T)}=\frac{1}{M^*(x)}-\int\frac{{\bf
p_F}}{p_F^2}\frac{\partial F({\bf p_F},{\bf p},x)}{\partial {\bf
p_F}}\delta n({\bf p},T)\frac{d{\bf p}}{(2\pi)^3}.
\end{equation}
We integrate the second term on the right hand side of Eq.
(\ref{M3}) over the angle variable and use the notation
\begin{equation}\label{F1}
    F_1(p_F,p,x)=Mp_F\int {\bf
p_F}\frac{\partial F({\bf p_F},{\bf p},x)}{\partial {\bf
p_F}}\frac{d\Omega}{(2\pi)^3},
\end{equation}
and substitute the variable $p$ by $z$, $z=(\varepsilon(p)-\mu)/T$.
Since in HF metals the  band is flat and narrow, we use the
approximation $(\varepsilon(p)-\mu)\simeq p_F(p-p_F)/M^*(T)$ and
taking into account Eqs. (\ref{M3}) and (\ref{F1}) finally obtain
\begin{eqnarray}
\nonumber \frac{M}{M^*(T)}&=&\frac{M}{M^*(x)}-
\alpha_1\int^{\infty}_{0}\frac{F_1(p_F,p_F(1 +\alpha_1 z),x)dz}{1+e^z} \\
&+\alpha_1&\int^{1/\alpha_1}_{0}F_1(p_F, p_F(1-\alpha_1
z),x)\frac{dz}{1+e^z}.\label{M4}
\end{eqnarray}
Here the factor $\alpha_1=TM^*(T)/p_F^2$. The Fermi momentum $p_F$
is defined from the relation $\varepsilon(p_F)=\mu$. We first
assume that $M^*(x)$ is finite and $\alpha_1\ll 1$. Then upon
omitting terms of the order of $\exp(-1/\alpha_1)$, we expand the
upper limit of the second integral on the right hand side of Eq.
(\ref{M4}) to $\infty$ and observe that the sum of the second and
third terms represents an even function of $\alpha_1$. These are
the typical expressions with Fermi-Dirac functions as integrands
and can be calculated using standard procedures \cite{lanl2}. We
conclude that at $T\ll T_F\sim p_F^2/M^*(x)$ the sum represents a
$T^2$-correction to $M^*(x)$ and the system demonstrates the LFL
behavior \cite{shag5}. When $x\to x_{FC}$ the effective mass
diverges and both $T_F\to0$ and $1/\alpha_1\to0$, while the
temperature interval over which the LFL behavior takes place is
vanishing \cite{shag5}. In that case, Eq. (\ref{M4}) becomes
homogeneous and the second integral on the right hand side can be
omitted. As a result, we can estimate that
\begin{equation}\label{MT}
    M^*(T)\propto \frac{1}{\sqrt{T}}.
\end{equation}
Equation (\ref{M4}) shows the universal power low behavior of the
effective mass which does not depend on the inter-particle
interaction. We illustrate this behavior by calculations using a
model functional \begin{eqnarray} \nonumber
E[n(p)]&=&\int\frac{{\bf p}^2}{2M}\frac{d{\bf
p}}{(2\pi)^3}+\frac{1}
{2}\int V({\bf p}_1-{\bf p}_2)\\
&\times&n({\bf p}_1)n({\bf p}_2) \frac{d{\bf p}_1d{\bf
p}_2}{(2\pi)^6}\label{MF},
\end{eqnarray} with the inter-particle interaction
\begin{equation}\label{MV}
 V({\bf p})=g_0\frac{\exp(-\beta_0|{\bf p}|)}{|{\bf
p}|}.\end{equation} We normalized the effective mass by $M$,
$M^*=M^*(T)/M$, temperature $T$ by the Fermi energy
$\varepsilon^0_F$, $T=T/\varepsilon^0_F$ and use the dimensionless
coupling constant $g=(g_0 M)/(2\pi^2)$ and $\beta=\beta_0 p_F$.
FCQPT takes place when the parameters reach their critical values,
$\beta=b_c$ and $g=g_c$, in our case $b_c=3$ and $g_c=6.7176$.
\begin{figure}[!ht]
\begin{center}
\includegraphics[width=0.47\textwidth]{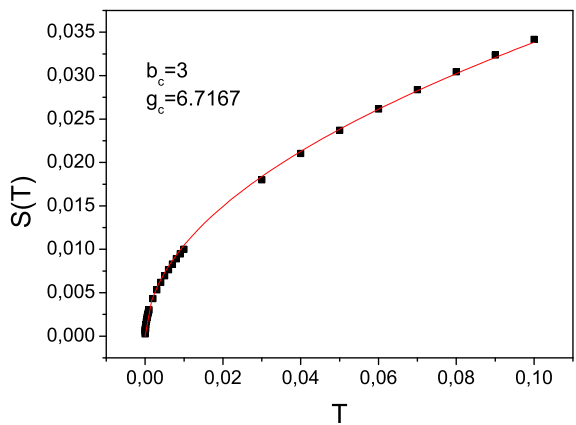}
\end{center}
\caption{Calculated entropy $S(T)$ as a function of temperature.
Continuous line represents $S(T)=c_1\sqrt{T}$, where $c_1$ is a
parameter. Solid squares are the results of calculations based on
functional (\ref{MF}).}\label{Fig2}
\end{figure}
In Fig.\ref{Fig2}, the evolution of the low temperature entropy is
shown. The calculated behavior of $S(T)/T\propto M^*(T)\propto
1/\sqrt{T}$ is in accord with Eq. (\ref{MT}).

Now consider the thermal expansion coefficient $\alpha(T)$  given
by \cite{lanl2} \begin{equation}
\alpha(T)=\frac13\left(\frac{\partial(\ln V)}{\partial T}\right)
_P=-\frac{1}{3V}\left(\frac{\partial(S/x)}{\partial
P}\right)_T,\label{ALP}
\end{equation}
Here, $P$ is the pressure and $V$ is the volume. The
compressibility $K(x)$ is not expected to be singular at FCQPT and
is approximately constant \cite{noz}. Inserting into Eq.
(\ref{ALP}) the entropy $S(T)\propto \sqrt{T}$, we find that
\begin{equation} \alpha(T)\simeq \frac{M^{*}T}{p_F^2K}\propto \sqrt{T}. \end{equation}
On the other hand, the
specific heat \begin{equation}
 C(T)=T\frac{\partial S(T)}{\partial
T}\propto \sqrt{T}.\label{ALP1}\end{equation} As a result, at
$T\to0$ the Gr\"uneisen ratio $\Gamma(T)$ tends to some constant
value rather than diverges as in the case when the electronic
system is on the ordered side of FCQPT \cite{zver,alp}
\begin{equation}
\Gamma(T)=\frac{\alpha(T)}{C(T)}=const\label{ALP2}.\end{equation}
Since the magnetic susceptibility $\chi(T)\propto M^*(T)$ and both
the Sommerfeld coefficient $C(T)/T\propto M^*(T)$ and
$\alpha(T)/T\propto M^*(T)$ we conclude that at $T\to0$
\begin{equation}\label{REL}
    \frac{C(T)}{T}\propto \chi(T)\propto\frac{\alpha(T)}{T}\propto
    M^*(T)\propto \frac{1}{\sqrt{T}}.
\end{equation}

\begin{figure} [! ht]
\begin{center}
\includegraphics [width=0.47\textwidth]{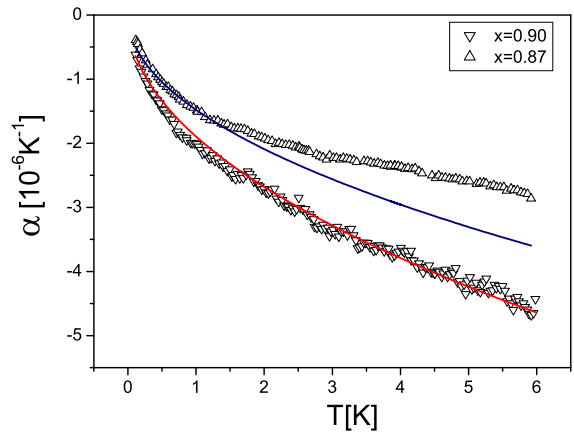}
\end{center}
\caption{The thermal expansion coefficient $\alpha(T)$ as a
function of temperature in the interval $\rm 100\, mK\leq T\leq 6\,
K$. Continuous curves are fits for $x= 0.87$ and $x=0.90$ data
\cite{sereni} based on Eq. (\ref{MT}) and represented by function
$\alpha(T)=c_1\sqrt{T}$ with $c_1$ being a fitting
parameter.}\label{AL}
\end{figure}
At this point, we consider how Eq. (\ref{REL}) and the behavior of
the effective mass given by Eq. (\ref{MT}) correspond to
experimental observations obtained on $\rm CePd_{1-x}Rh_x$.
Measurements of the thermal expansion coefficient $\alpha(T)$ on
$\rm CePd_{1-x}Rh_x$ with $x=0.87$ and $x=0.90$ \cite{sereni} are
shown in Fig. \ref{AL}. It is seen that the approximation
$\alpha(T)=c_1\sqrt{T}$ for composition $x=0.90$ is in good
agreement with facts over two orders of magnitude in the
temperature range from 6 K down to 100 mK, and measurements on
${\rm CeNi_2Ge_2}$ \cite{geg1} and $\rm CeIn_{3-x}Sn_x$ \cite{kuch}
demonstrate the same behavior. While $\rm CePd_{1-x}Rh_x$ is a
three dimensional ferromagnet \cite{sereni,pikul},  $\rm
CeNi_2Ge_2$ exhibits a paramagnetic ground state \cite{geg1} and
$\rm CeIn_{3-x}Sn_x$ is antiferromagnetic cubic metal \cite{kuch}.
We conclude that the observed uniform behavior of the thermal
expansion coefficient of these metals is determined by
quasiparticles and FCQPT rather than by different magnetic quantum
critical points and corresponding fluctuations. Measurements of the
specific heat $C(T)$ on $\rm CePd_{1-x}Rh_x$ with $x=0.87$ and
$x=0.90$  show a power law $T$ dependence. These are described by a
$C(T)/T=AT^{-q}$ formula with the exponent $q\simeq 0.5-0.4$ and
$A$ is a constant, around that concentration $C(T)/T$ and $\chi(T)$
coincide in their $T^{-q}$ temperature dependence
\cite{sereni,pikul}. We conclude that the results given by  Eq.
(\ref{REL}) agree with facts.

\begin{figure} [! ht]
\begin{center}
\includegraphics [width=0.47\textwidth]{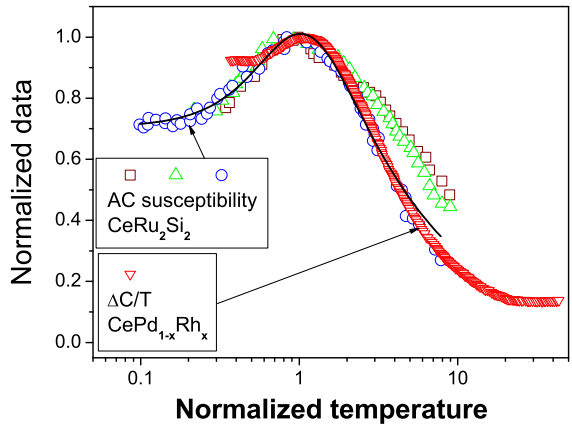}
\end{center}
\caption{Normalized magnetic susceptibility
$\chi_{AC}(T,B)/\chi_{AC}(T_M,B)$ for $\rm CeRu_2Si_2$ in magnetic
fields 0.20 mT (squares), 0.39 mT (upright triangles) and 0.94 mT
(circles) is plotted against normalized temperature $T/T_M$
\cite{takah}.  The susceptibility reaches its maximum value
$\chi_{AC}(T_M,B)$ at $T=T_M$. Normalized 4f electron contribution
$(\Delta C(T)/T)/(\Delta C(T_M)/T_M)$  to the specific heat of $\rm
CePd_{1-x}Rh_x$ with $x = 0.80$ versus normalized temperature
$T/T_M$ is shown by downright triangles  \cite{pikul}. Here $T_M$
is the temperature at the peak of $\Delta C(T)/T$. The solid curve
traces the universal behavior of the normalized effective mass
determined by Eq. (\ref{UN2}).}\label{UB}
\end{figure}

Consider the case when the concentration $x$ deviates from the
critical value $x_{FC}$ so that the system is moved to the
disordered side of FCQPT and temperature $T_F$ becomes finite. As a
result, at $T\ll T_F$ the system exhibits the LFL behavior with the
effective mass being approximately constant, $M^*(T)\simeq
M^*(x)+a_1(T/T_F)^2$, where the term $a_1(T/T_F)^2$ represents
correction to $M^*(x)$. If $x=x_{FC}$ then the application of
magnetic field making $T_F$ finite restores the LFL behavior with
the effective mass depending on $B$ as \cite{smag,ckhz}
\begin{equation}\label{B32}
    M^*(B)\propto (B-B_{c0})^{-2/3},
\end{equation}
where $B_{c0}$ is the critical magnetic field which drives both a
HF metal to its magnetic field tuned QCP and the corresponding
N\'eel temperature toward $T=0$. In some cases $B_{c0}=0$, for
example, the HF metal CeRu$_2$Si$_2$ is characterized by $B_{c0}=0$
and shows neither evidence of the magnetic ordering,
superconductivity nor the LFL behavior down to the lowest
temperatures \cite{takah}.

At $T\sim T_F$, the effective mass depends mainly on temperature
\cite{ckhz,shag5}
\begin{equation}\label{T32}
    M^*(T)\propto T^{-2/3}.
\end{equation}
Then, at elevated temperatures $T\gg T_F$, the behavior of the
effective mass is given by Eq. (\ref{MT}) \cite{shag5}. Therefore
at $T\lesssim T_F$ the behavior of the effective mass can be
described by a simple function \cite{s_univ}
\begin{equation}
\frac{M^*(B,T)}{M^*(B)}\approx\frac{1+c_2y^2}{1+c_3y^{8/3}},
\label{UN1}\end{equation} which represents an approximation to
solutions of Eq. (\ref{M1}) that agrees with Eqs. (\ref{B32}) and
(\ref{T32}). Here $y=(T/(B-B_{c0}))$, $c_2$ and $c_3$ are fitting
parameters. Since the effective mass reaches its maximum value
$M^*_M$ at some $y=y_M$ \cite{shag5,ckhz} we define a normalized
effective mass as $M^*_N(T,B)=M^*(T,B)/M^*_M$. Taking into account
Eq. (\ref{UN1}) and introducing the variable $z=y/y_M$ we obtain
the function
\begin{equation}
M^*_N(z)\approx\frac{1}{M^*_M}\frac{1+c_2z^2}{1+c_3z^{8/3}},
\label{UN2}\end{equation} which describes a universal behavior of
the effective mass $M^*_N(z)$. In the case of finite $M^*(x)$, Eq.
(\ref{UN2}) is valid at $T\sim T_F$ if $M^*(T,B)/M^*(x)\ll 1$
because the term $1/M^*(x)$ on the right hand side of Eq.
(\ref{M3}) being a small correction to the effective mass can be
omitted \cite{samm}. It is seen from Eq. (\ref{UN2}) that
$M^*_N(z)$ reaches its maximum value at $z=1$, $M^*_N(z=1)=1$.

The effective mass $M^*(T,B)$ can be measured in experiments on HF
metals. For example, as it follows from Eq. (\ref{REL})
$M^*(T,B)\propto C/T$ and $M^*(T,B)\propto \chi_{AC}(T)$ where
$\chi_{AC}(T)$ is the magnetic susceptibility. If the corresponding
measurements are carried out at fixed value of magnetic field $B$
(or at fixed value of the concentration $x$ and $B=0$) then as it
follows from Eq. (\ref{UN1}) the effective mass reaches the maximum
at some temperature $T_M$. Upon normalizing both the effective mass
by its peak height at each field $B$  and the temperature by $T_M$,
we observe that all the curves should demonstrate a scaling and
collapse on the single curve given by Eq. (\ref{UN2}).

As shown in Fig. \ref{UB}, the  behavior of the normalized
susceptibility
$\chi_{AC}^N(z)=\chi_{AC}(T/T_M,B)/\chi_{AC}(1,B)=M^*_N(z)$
obtained in measurements on the HF paramagnetic $\rm CeRu_2Si_2$
\cite{takah} is in accord with the approximation given by Eq.
(\ref{UN2}). Since the crossover temperature $T^*$ from the regime
given by Eq. (\ref{T32}) to the regime given by Eq. (\ref{MT}) is
proportional to the magnetic field, $T^*\propto B$
\cite{shag5,s_univ}, we expect that the temperature range over
which the scaling takes place shrinks when the applied magnetic
field $B$ is diminished. It is seen from Fig. \ref{UB} that the
deviation of the data corresponding to the smallest value of the
magnetic field $B=
 0.20$ mT (shown by squares) is largest at the elevated normalized
temperature, while the slope of this curve tends to that of curve
described by Eq. (\ref{MT}). At small temperatures as seen from
Fig. \ref{UB}, both the curve given by Eq. (\ref{UN2}) and the
effective mass determined by Eq. (\ref{B32}) agree perfectly with
facts collected in measurements on $\rm CeRu_2Si_2$ whose
electronic system is placed at FCQPT \cite{s_univ}. As to the
normalized 4f contribution $(\Delta C(T)/T)/(\Delta
C(T_M)/T_M)=M^*_N(T)$  (shown by downright triangles in Fig.
\ref{UB}) to the specific heat of $\rm CePd_{1-x}Rh_x$ with $x =
0.80$ \cite{pikul}, the scaling takes place up to relatively large
temperatures because the deflection of the $x=0.8$ from the
critical concentration $x_{FC}\simeq 0.9$ is big, while $T^*\propto
B\propto |x_{FC}-x|$ \cite{samm}. As a result, at diminishing
temperatures the scaling is ceased at relatively high temperatures
as soon as the LFL behavior sets in.

In summary, we have shown that the NFL behavior of the thermal
expansion coefficient $\alpha(T)$ observed in $\rm CePd_{1-x}Rh_x$
at the critical concentration $x_c\simeq 0.9$ coincides with that
of $\alpha(T)$ observed in both $\rm CeNi_2Ge_2$ exhibiting the
paramagnetic ground state and antiferromagnetic cubic HF metal $\rm
CeIn_{3-x}Sn_x$. We have also shown that the behavior of the
normalized effective mass $M^*_N$ observed in $\rm CePd_{1-x}Rh_x$
at $x\simeq 0.8$ agrees with that of $M^*_N$ observed under the
application of magnetic field in paramagnetic $\rm CeRu_2Si_2$ and
concluded that these alloys exhibit the universal NFL thermodynamic
behavior at its QCP. The outlined behavior is independent of the
characteristic features of the given alloys while numerous CQPs
assumed to be responsible for the NFL behavior of different HF
metals can be substituted by the only QCP related to FCQPT.

This work was supported in part by RFBR, project No. 05-02-16085.


\begin{thebibliography}{199}

\bibitem{voj} M. Vojta, Rep. Prog. Phys. {\bf 66}, 2069 (2003).

\bibitem{loh} H.v. L\"ohneysen, A. Rosch, M. Vojta, and P. W\"olfle,
cond-mat/0606317.

\bibitem{ste} G.R. Stewart, Rev. Mod. Phys. {\bf 73}, 797 (2001).

\bibitem{sereni} J.S. Sereni {\it et. al.,} cond-mat/0602588.

\bibitem{pikul} A.P. Pikul {\it et. al.,} J. Phys.: Condens. Matter {\bf
18}, L535 (2006).

\bibitem{kuch} R. K\"uchler,  {\it et al.,} Phys. Rev. Lett. {\bf 96}, 256403 (2006).

\bibitem{geg1}  R. K\"uchler {\it et. al.,} Phys. Rev. Lett. {\bf 91},
066405 (2003).

\bibitem{kir} T.R. Kirkpatrick and D. Belitz, Phys. Rev. B {\bf 67}, 024419 (2003).

\bibitem{shag} V.R. Shaginyan JETP Lett. {\bf 79}, 286 {2004}.

\bibitem{pag} J. Paglione  {\it et al.,} Phys. Rev. Lett. {\bf 97} (2006) 106606.

\bibitem{shag1} V.R. Shaginyan, A.Z.  Msezane, V.A. Stephanovich, and E.V.
Kirichenko, Europhys. Lett. {\bf 76}, 898 (2006).

\bibitem{ams} M.Ya. Amusia and V.R. Shaginyan, Phys. Rev. B {\bf 63}, 224507 (2001).

\bibitem{ams1} M.Ya. Amusia and V.R. Shaginyan,  JETP Lett. {\bf
73}, 232 (2001).

\bibitem{land} E.M. Lifshitz, L.P. Pitaevskii {\it Statistical
Physics} Part 2, Butterworth-Heinemann, Oxford (1999).

\bibitem{pfit} M. Pfitzner and P. W\"olfle P. Phys. Rev.
B {\bf 33}, 2003 (1986).

\bibitem{shag_fc} V.R. Shaginyan, JETP Lett. {\bf 77}, 99 (2003).

\bibitem{lanl2}  E.M. Lifshitz and L.P. Pitaevskii,
{\it Statistical Physics} Part 1, Butterworth-Heinemann, Oxford
(2000).

\bibitem{shag5}  V.R. Shaginyan, JETP Lett. {\bf 80}, 263  (2004).

\bibitem{noz}  P. Nozi\`eres, J. Phys. I (France) {\bf 2}, 443  (1992).

\bibitem{zver} M.V. Zverev, V.A. Khodel, and V.R. Shaginyan,
JETP Lett. {\bf 65}, 863 (1997).

\bibitem{alp} M.Ya. Amusia, A.Z. Msezane, and V.R. Shaginyan,
Phys. Lett. A {\bf 320}, 459 (2004).

\bibitem{smag} V.R. Shaginyan, JETP Lett. {\bf 77}, 178 (2003).

\bibitem{ckhz} J.W. Clark, V.A. Khodel, and M.V. Zverev,
Phys. Rev B {\bf 71}, 012401 (2005).

\bibitem{takah} D. Takahashi {\it et al.,}
Phys. Rev. B {\bf 67}, R180407 (2003).

\bibitem{s_univ} V.R. Shaginyan, JETP Lett.  {\bf 79}, 286 (2004).

\bibitem{samm} V.R. Shaginyan, M.Ya. Amusia, and A.Z. Msezane, Phys.  Lett.  A {\bf 338}, 393 (2005).

\end{thebibliography}
\end{document}